\documentclass[fleqn,usenatbib]{mnras}

\usepackage[T1]{fontenc}
\usepackage{ae,aecompl}
\usepackage{afterpage}

\usepackage{graphicx}	
\usepackage{amsmath}	
\usepackage{amssymb}	

\newcommand{\be}{\begin{equation}}
\newcommand{\ee}{\end{equation}}
\newcommand{\ms}{_{\odot}}

\usepackage[normalem]{ulem}
\usepackage{color}

\title[GW inspirals inside supermassive stars]{Probing supermassive stars and massive black hole seeds through gravitational wave inspirals}

\author[Raveh, Ginat, Perets \& Woods]{
Yael Raveh,\thanks{E-mail: yael.raveh@campus.technion.ac.il}
Yonadav Barry Ginat,
Hagai B. Perets
and Tyrone E. Woods
\\
Faculty of Physics, Technion -- Israel Institute of Technology, Haifa, 3200003, Israel\\
National Research Council of Canada, Herzberg Astronomy \& Astrophysics Research Centre, 5071 West Saanich Road,\\ Victoria, BC V9E 2E7, Canada
}

\date{Accepted XXX. Received YYY; in original form ZZZ}

\pubyear{2020}

\begin{document}
\label{firstpage}
\pagerange{\pageref{firstpage}--\pageref{lastpage}}
\maketitle

\begin{abstract}
We propose a novel source of gravitational wave emission: the inspirals of compact fragments inside primordial supermassive stars (SMSs).
Such systems are thought to be an essential channel in the as-yet little understood formation of supermassive black holes (SMBHs). One model suggests that high accretion rates of $0.1$-1 M$_\odot$/yr attainable in atomically-cooled primordial halos can lead to the formation of a nuclear-burning SMS. This will ultimately undergo collapse through a relativistic instability, leaving a massive BH remnant. Recent simulations suggest that supermassive stars rarely form in isolation, and that companion stars and even black holes formed may be captured/accreted and inspiral to the SMS core due to gas dynamical friction.
Here, we explore the GW emission produced from such inspirals, which could probe the formation and evolution of SMS and seeds of the first supermassive black holes. 
We use a semi-analytic gas-dynamical friction model of the inspirals in the SMS to characterize their properties.
We find such sources could potentially be observable by upcoming space-born GW-detectors at their formation redshifts with the benefit of gravitational lensing. Mergers within closely-related quasi-stars may produce a much stronger signal, though disambiguating such events from other high-z events may prove challenging.
\end{abstract}


\section{Introduction}
Most galaxies are thought to harbor a massive black hole at their center (e.g., \citealp{2013ARA&A..51..511K,2005SSRv..116..523F}). However, the origin of such massive black holes (SMBHs) and their evolution and growth are still little understood.  The presence of billion-solar-mass quasars at very high redshifts ($\sim 7$), indicates that at least some SMBHs were formed very early in the history of the Universe.

SMBHs may have evolved from initial ``seeds'' via a number of proposed channels \citep{2010A&ARv..18..279V, 2019PASA...36...27W, 2020ARA&A..58...27I}, with masses in the range $10^2-10^5 M\ms$ and forming at redshifts $10\lesssim z\lesssim 20$. These seeds could be provided by the remnants of 
atypical Pop III stars within atomically-cooled gaseous halos 
(e.g., \citealp{2008MNRAS.391.1961D}), in which ``normal'' Pop III star formation is suppressed and infall rates of up to $\sim 0.1-1.0$\,M$\ms\,yr^{-1}$ are possible. Numerical simulations have shown such high accretion rates lead to the formation of a nuclear burning, super-massive star (SMS), which later undergoes collapse through a relativistic instability, leaving a massive BH remnant (e.g, \citealp{2013ApJ...778..178H,2017ApJ...842L...6W, 2018MNRAS.474.2757H}). 
Another possibility is that the seeds are formed directly from the collapse of dense gas clouds in the inner regions of gaseous proto-galaxies \citep{maggiore2008gravitational}, potentially allowing for even higher infall rates \citep{2010Natur.466.1082M}.

Seed BHs are necessarily a transient population of objects, and inferring their initial mass function and spin distribution from observations is possible only if they can be detected either through electromagnetic or GW observations at very high $z$, as high as $\sim 20$. BHs of any mass which are bound in binaries are loud sources of GWs at the time of their merging, therefore unveiling the seeds of SMBHs via their GW emission at coalescence \citep{2007MNRAS.377.1711S} may provide unique and invaluable information on BH genesis and evolution, and is probably one of the best ways to discriminate among formation mechanisms.

The structure of such supermassive objects, their response to rapid accretion, and the nature of their surrounding environments, have been studied intensively in the last few years \citep[see][for a review]{2019PASA...36...27W}. Mounting evidence from recent numerical simulations of primordial halos has shown that SMSs are not typically born in isolation \citep[e.g.,][]{2020ApJ...892L...4L}, and that they may interact with companion objects both before and after they collapse to black holes \citep{2021arXiv210208963W}. \cite{2015MNRAS.452..755S} crudely investigated the response of SMSs to episodic accretion, stemming from 
fragments, which were formed via gravitational instability, that migrate inward 
and accrete onto the star, while 
\cite{2019arXiv190910517T} suggested that subsequent frequent capture and accretion of stars onto the SMS could inhibit the collapse of its core, and give rise to a larger scale collapse which leaves a more massive remnant behind, with masses as high as $\sim 10^5 - 10^6M\ms$. 
Three-dimensional radiation-hydrodynamic simulations of primordial, atomic-cooling halos by e.g., \citet{2019Natur.566...85W}, \citet{2020ApJ...892L...4L}, and \citet{2020arXiv201211612P} have shown that the formation of multiple SMSs, as well as smaller stellar-mass fragments, may be ubiquitous aspects of supermassive star formation in such environments. This suggests that interactions between supermassive stars and other massive and supermassive objects may be common in the early Universe, and their detection may provide an invaluable probe of the formation of SMBHs.

Third generation GW observatories hold substantial promise for extending our understanding of binaries with total mass of $10^4 - 10^6M\ms$ \citep{2017PhRvD..95j3012B}. Since SMSs and the compact remnants they form are in this range, such GW-detectors might be sensitive to the inspiral of compact objects and stars inside SMSs, and in particular in cases where binary SMSs may form (e.g., \citealp{2018MNRAS.475.4104C}) and later inspiral and merge \citep{2018MNRAS.479L..23H}. Depending on the stellar evolution of supermassive binaries \citep{2021arXiv210208963W} and the initial separation of their compact remnants, these objects will merge and can, in principle, be observed using future space-based GW detectors \citep{2020ApJ...892L...4L}. Here we study the gravitational radiation from such mergers of compact objects (COs) or massive stars inspiralling onto a ``seed''-like object, through its gaseous halo. We follow the inspiral due to gas-dynamical friction and characterise the expected GW signals from such inspirals, using a similar approach to our previous study of the inspiral and GW characteristics of COs inspiralling inside a regular binary common-envelope \citep{2019arXiv190311072G}.

In section~\ref{sec:seed} we depict the density and temperature profiles used for the SMSs in our calculations. Sec.~\ref{sec:method} outlines our methods for characterizing the inspirals and their GW signatures. In sec.~\ref{sec:results} we apply our model to a grid of COs/stars inspiralling inside SMSs at high redshifts ($z \sim 6$)
and present our results, which we then discuss and summarize in Sec.~\ref{sec:sum}.

\section{Methods} \label{sec:method}
In order to study inspirals inside SMSs, we first describe the SMS models we use in section \ref{sec:seed}, we then review our gas-dynamical friction inspiral model (\ref{sec:model}), and then describe the GW signature characterization (\ref{sec:waveform}).

We used density and temperature profiles of SMSs from both numerical simulations and semi-analytical calculations, described in section~\ref{sec:seed}. These profiles were then implemented in a code that solves the equation of motion given an arbitrary companion, using Ostriker's model for gas dynamical friction (described in section~\ref{sec:model}), and calculates the characteristic strain and expected SNR. Note that we neglect the back-reaction of
the motion of the system on the SMS itself. 
The waveforms emitted by the binary are calculated using the derivation presented in section~\ref{sec:waveform}, and detectability prospects are estimated using the method described in section~\ref{sec:snr}.

\subsection{Profile of the accreting SMS} \label{sec:seed}

\begin{figure}
\includegraphics[scale=0.4,trim={0cm 5.55cm 0cm 6.3cm},clip]{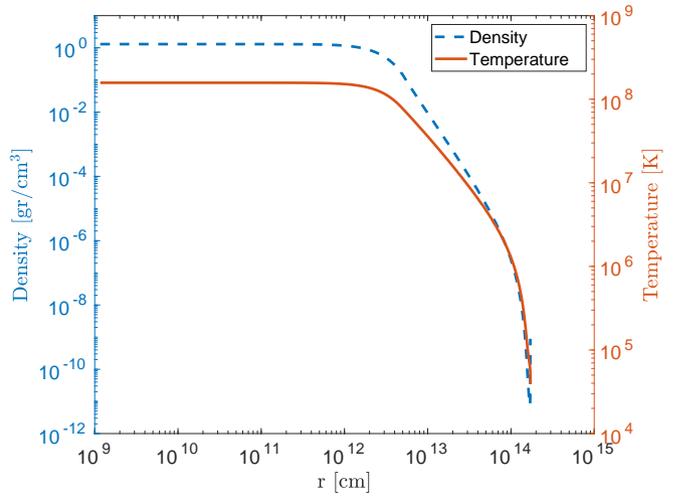}
\caption{Radial profiles for the density and temperature of the rapidly-accreting supermassive star once it has reached approximately $2.09\times 10^5 M\ms$.
} \label{fig:profile}
\end{figure}
The maximum mass attainable by rapidly-accreting SMSs depends on the specific formation channel being invoked, as well as details of the numerical treatment of the relevant physics in the case for differing simulations \citep{2019PASA...36...27W}. For all models however, a limit of several $10^5 M\ms$ (or less) is broadly applicable for a range in accretion rates spanning the two orders of magnitude found in simulations of atomic-cooling halos. Models of SMSs accreting at $100 - 1000 M\ms yr^{-1}$  in the even more extreme environments thought to be viable in the merger of some gas-rich proto-galaxies \citep{2010Natur.466.1082M} suggest that this limit remains always below $10^6 M\ms$ even for such extreme rates \citep{2020arXiv201008229H,2020A&A...644A.154H}.

We use two SMS models; the first is taken from \cite{2017ApJ...842L...6W}, where the birth, evolution, and collapse of accreting, non-rotating SMSs were studied under varied accretion rates using the 1D Lagrangian hydrodynamics and stellar evolution code named \emph{Kepler} \citep{2002RvMP...74.1015W}. 
Fig.~\ref{fig:profile} shows radial profiles for the density and temperature of our example rapidly-accreting SMS, given an accretion rate of $0.316 M\ms\,yr^{-1}$, once it has reached approximately $2.09\times 10^5 M\ms$, i.e. a few years before its collapse due to the onset of the post-Newtonian instability. Additionally, we use an analytic hylotropic model of a $M_{SMS}=10^7 M\ms$ Pop III SMS, with a $M=10^4 M\ms$ core, adapted from \citet{2020A&A...644A.154H} to model an extremely supermassive object formed in the extreme accretion rates which may be possible in the merger of some gas-rich galaxies.

\subsection{Model description} 
\label{sec:model}
Much like in \cite{2019arXiv190311072G} (cf. for more details), where the GW signal from inspirals of COs through a gaseous common envelope of an aged stellar companion was studied, we model the system as an effective single body, whose position $\textbf{r}$ is the displacement vector between the companion and the core of the evolved star, and neglect any back-reaction of the companion on the SMS. 
This approximation is reasonable given that the SMS is significantly more massive than the companion; indeed, we stop the evolution once the mass inside the CO instantaneous orbit becomes comparable to its own mass. 
We note that in principle, the inspiralling object may disrupt the inner region/core of the SMS to eventually inspiral to the center, and may give rise to higher frequency GW signal. Such late evolution, however, can not be modelled through our semi-analytic approach for gas-dynamical friction, but could be studied in the future through hydrodynamical simulations. These are beyond the scope of the current study.

Let $M_{env}(r)$ be the mass inside a sphere of radius $r$ in the SMS (excluding the core), $M = M_{core}$ -- the SMS core mass, $m$ -- the companion mass, and $\mu$ -- their reduced
mass. The equations of motion are then
\be \mu \ddot{\textbf{r}}=-\frac{G\mu (M+m)}{r^3}\textbf{r}-\frac{G\mu M_{env}(r)}{r^3}\textbf{r}-F(\textbf{r},\textbf{v})\textbf{v}+P.N., \label{eq:EOM} \ee
where `P.N.' refers to post-Newtonian terms\footnote{PN corrections of up to 2.5PN are included, as given by, e.g., \cite{1990PhRvD..42.1123L}, but in reality their effect on the entire evolution is minuscule.}. 

The function $F$ describes gas dynamical friction between the envelope and the CO; we use the model of \cite{1999ApJ...513..252O}:
\be F(\textbf{r}, \textbf{v})=\frac{2\pi G^2m\mu \rho(r)}{v^3}\begin{cases}
\ln\left(\frac{1+\mathcal{M}}{1-\mathcal{M}}e^{-2\mathcal{M}}\right), & \mathcal{M}<1\\
\ln\left(\Lambda^{2}-\frac{\Lambda^{2}}{\mathcal{M}^{2}}\right), & \mathcal{M}>1.
\end{cases}, \ee
where $\mathcal{M}=v/c_{s}$ is the Mach number and $c_{s}$ is the local sound speed.\footnote{The Coulomb logarithm $\ln\Lambda$ is $\Lambda=b_{\max}/b_{\min}$ \citep{binney2011galactic}, where $b_{\min}=\max\left\lbrace Gm/v^{2},r_{\textrm{coll}}\right\rbrace $, $r_{\textrm{coll}}$ is the radius where the CO and the core collide; we take $b_{max} = 2r$, as in \cite{2007ApJ...665..432K} rather than $b_{max}=R_{SMS}$ as one might na\"{i}vely expect, because the density outside $r$ is much smaller than the density inside $r$.}

\subsection{The GW waveforms} \label{sec:waveform}
We use the standard Einstein quadrupole radiation formula \citep{maggiore2008gravitational},
\be h^{TT}_{ij}(\textbf{x},t)=\frac{2G}{c^4D}\ddot{Q}^{TT}_{ij}(t-D/c), \ee
where $D$ is the distance to the source, and $Q^{TT}$ denotes the trace free part of the mass-quadrupole moment. 

\subsubsection{Detectability prospects} \label{sec:snr}
The characteristic strain quantifies the detectors' sensitivity to the GW signal discussed in this paper. If $S_n(f)$ is the noise power spectrum density of a detector, and $h$ is the physical signal (without noise) then the SNR is given by
\be \left( \frac{S}{N}\right) ^2=4\int_0^{\infty}{\frac{\vert\tilde{h}(f)\vert^2}{S_n(f)}df}\equiv \int_0^{\infty}{\frac{h^2_c(f)}{h^2_n(f)}df}, \ee
where $\tilde{h}(f)=\int_{-\infty}^{\infty}{e^{-2\pi ift}h(t)dt}$ \citep{2015CQGra..32a5014M}. 
In practice these are calculated using a discrete fast Fourier transform, after applying a Tukey window \citep{2016PhRvD..93l2003A}. The observed frequency is related to the source frequency by $f_{obs}=f_{src}/(1+z)$ with $z$ being the source's redshift. Luminosity distances are computed by assuming the following cosmology \citep{Planck2018}: $H_0=67.4\,$km~s$^{-1}$~Mpc$^{-1}$, $\Omega_m=0.32$, $\Omega_\Lambda = 0.68$. 

\section{Results} \label{sec:results} 
\begin{figure}
\includegraphics[scale=0.3,trim={3.6cm 6cm 4cm 5.9cm},clip]{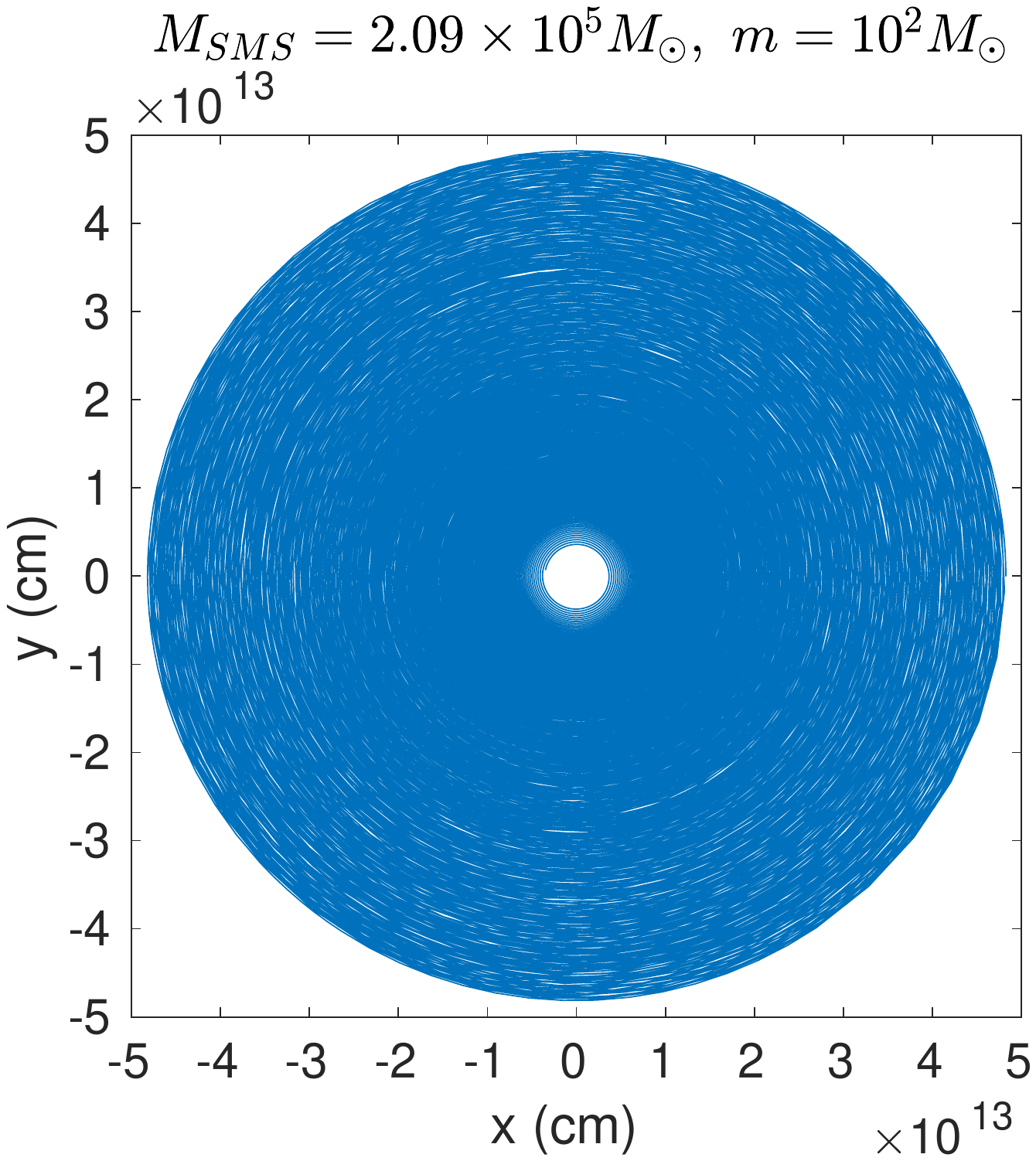} \includegraphics[scale=0.3,trim={3.6cm 6cm 4cm 5.9cm},clip]{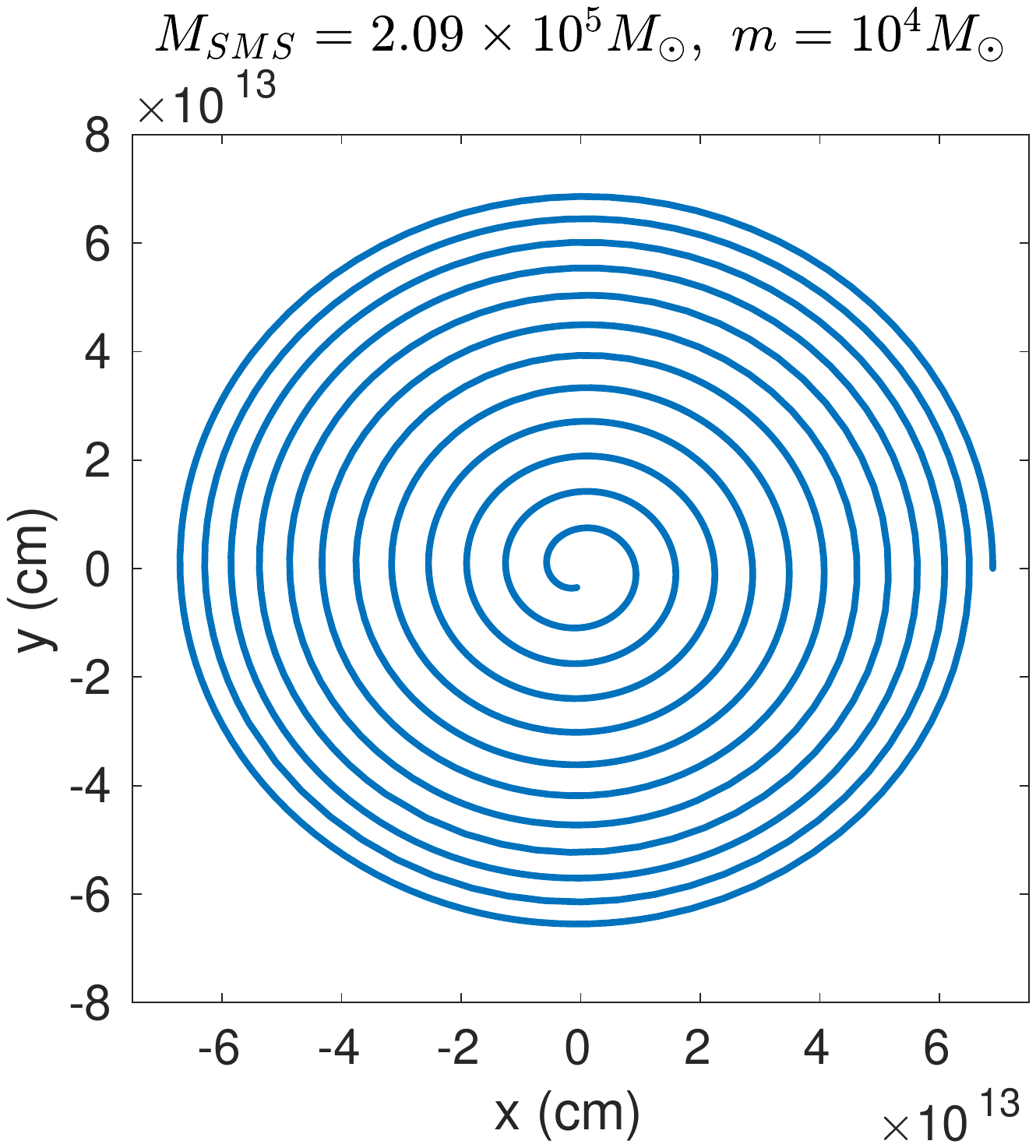}
\includegraphics[scale=0.3,trim={3.6cm 6.7cm 4cm 5.5cm},clip]{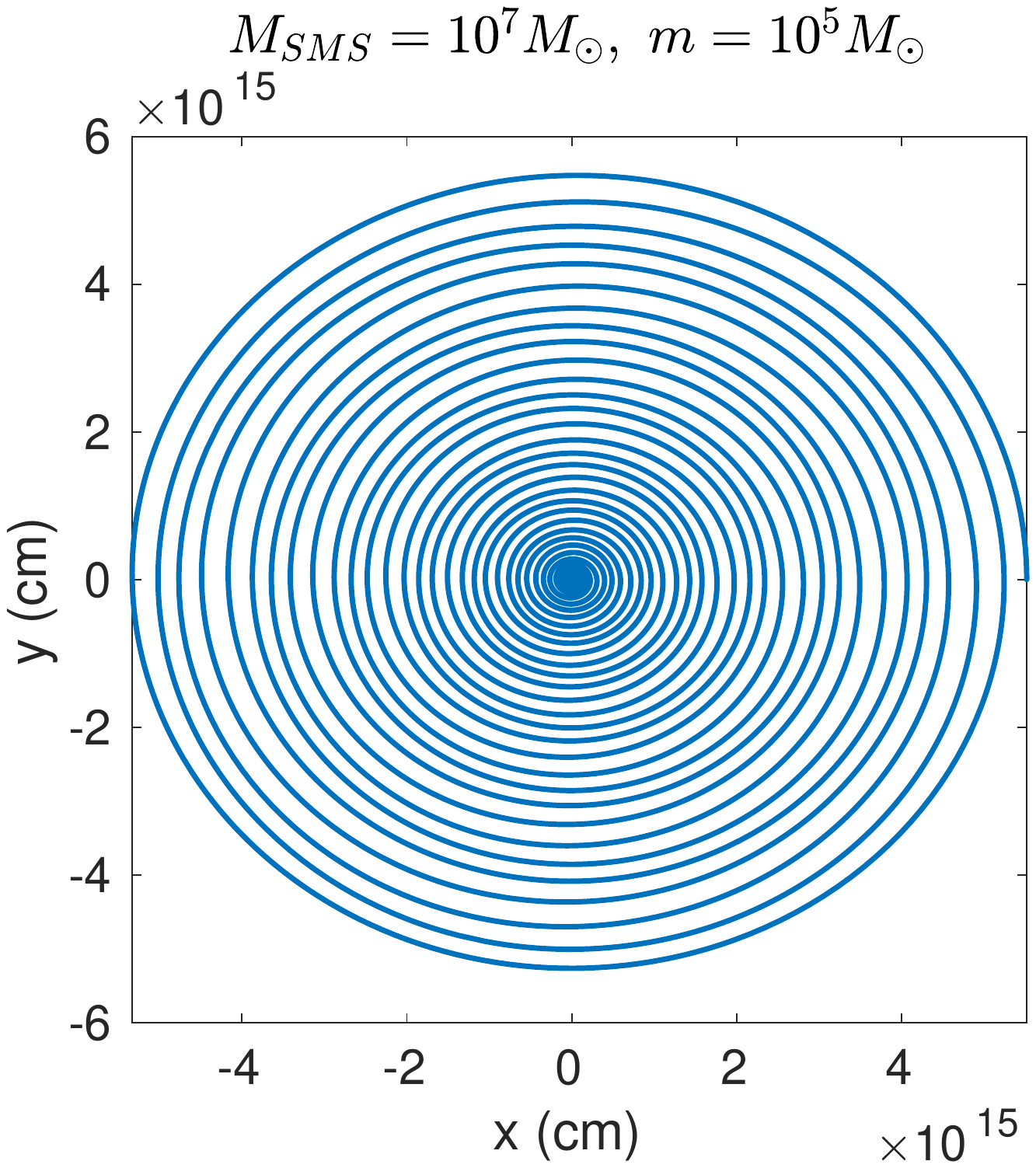} \includegraphics[scale=0.3,trim={3.5cm 6.7cm 3.7cm 5.5cm},clip]{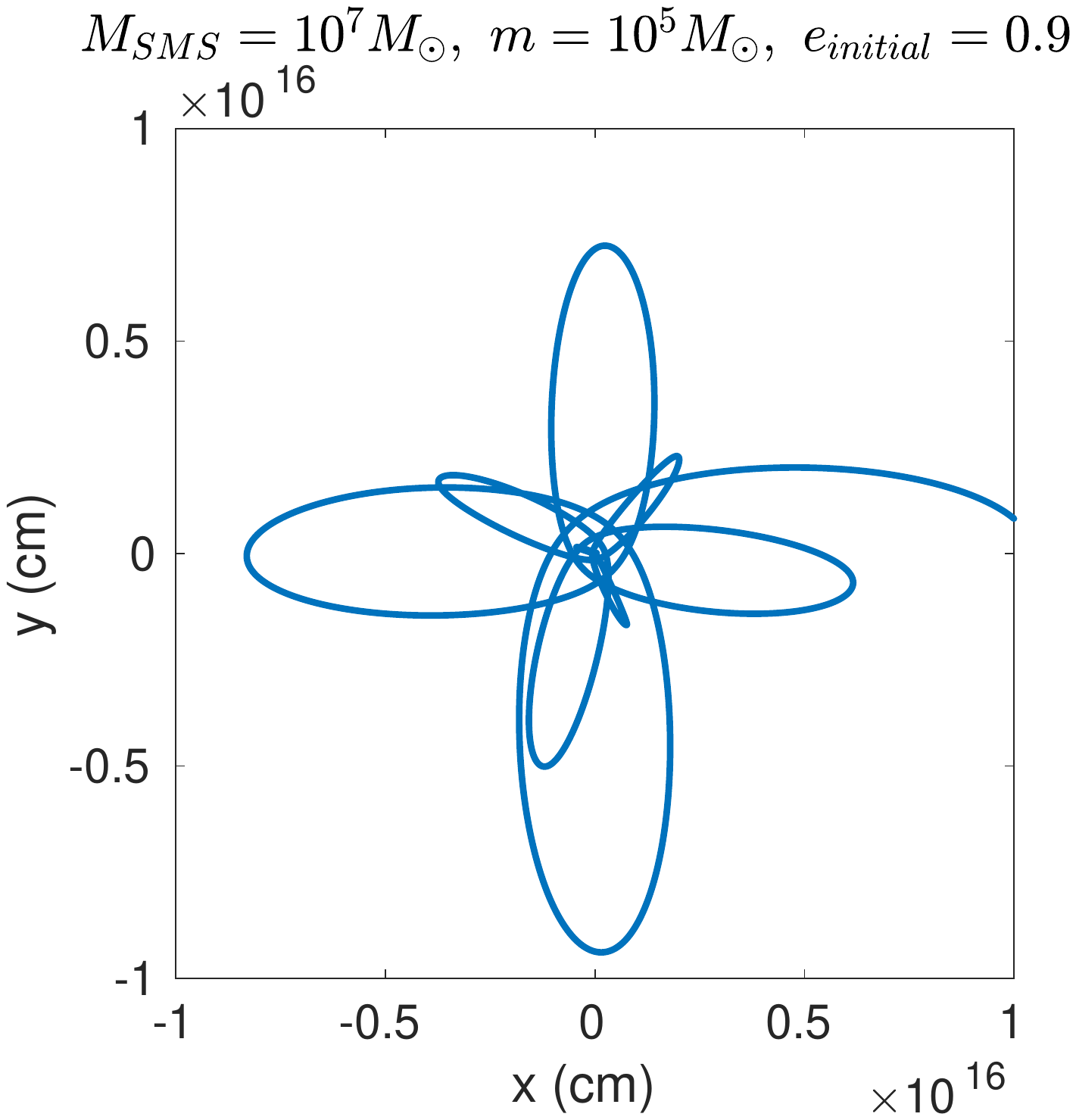}
\caption{Separation between the companion and the SMS's core. Top-left (top-right): an SMS of total mass $2.09\times 10^5 M\ms$ with a companion of mass  $m=10^2 M\ms$ ($m=10^4 M\ms$). Bottom: an SMS of total mass $10^7 M\ms$ with a companion of mass $m=10^5 M\ms$.
} \label{fig:orb}
\end{figure}

In order to calculate the expected GW signature from the inspiral, we integrated equation \eqref{eq:EOM} numerically using a Runge-Kutta integrator. 
Examples of such orbits computed as described above are shown in Fig.~\ref{fig:orb}. The bottom right figure shows an example eccentric orbit assuming the companion is an SMS in a fragmented accretion disk at the center of an atomically-cooled halo \citep{2020ApJ...892L...4L} that exhibits morphology of fragments with highly eccentric orbits \citep{2020arXiv201211612P}. 
\begin{figure}
\includegraphics[scale=0.245,trim={2.25cm 6.65cm 1.63cm 6.8cm},clip]{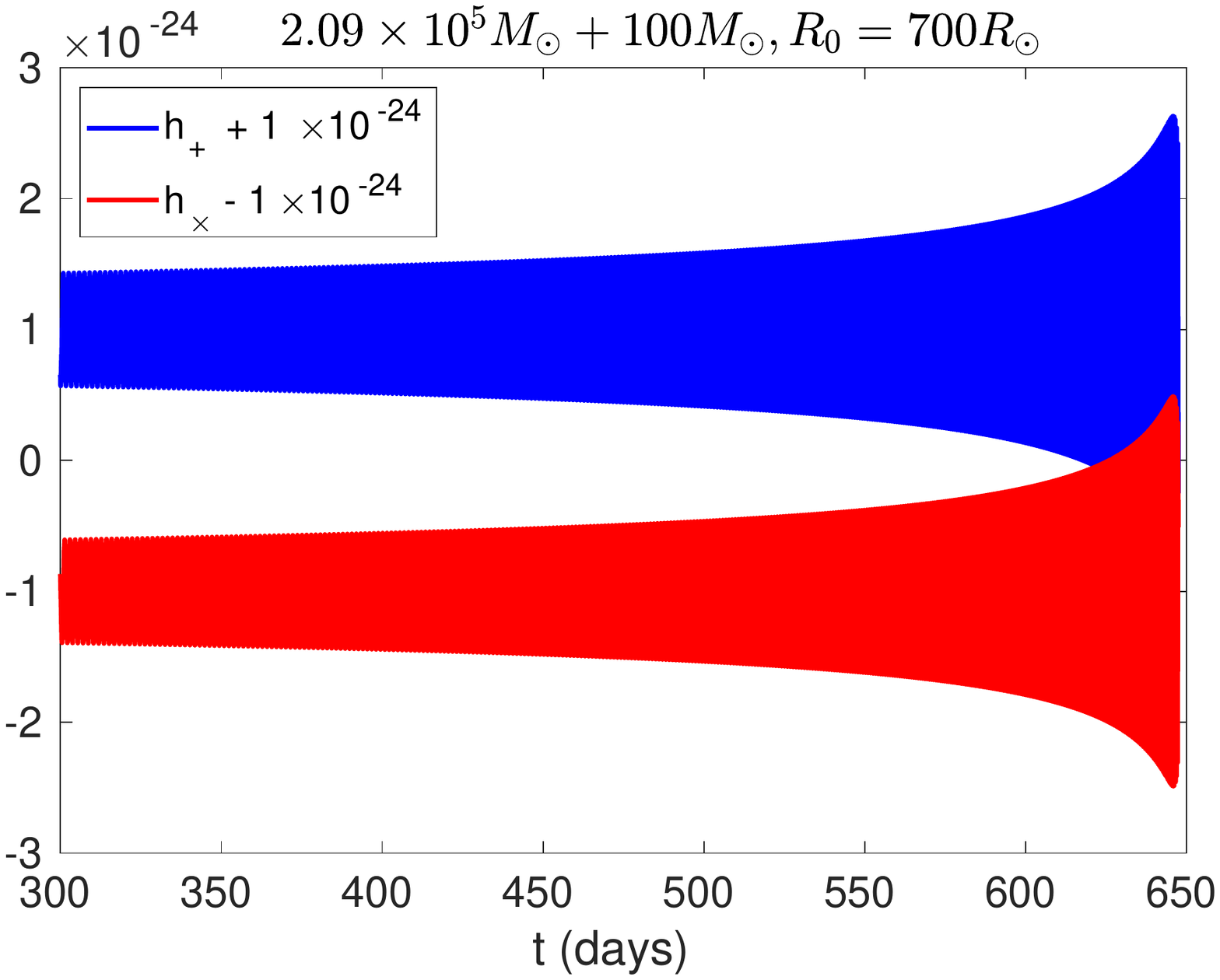} \includegraphics[scale=0.247,trim={2.25cm 6.0cm 1.8cm 6.3cm},clip]{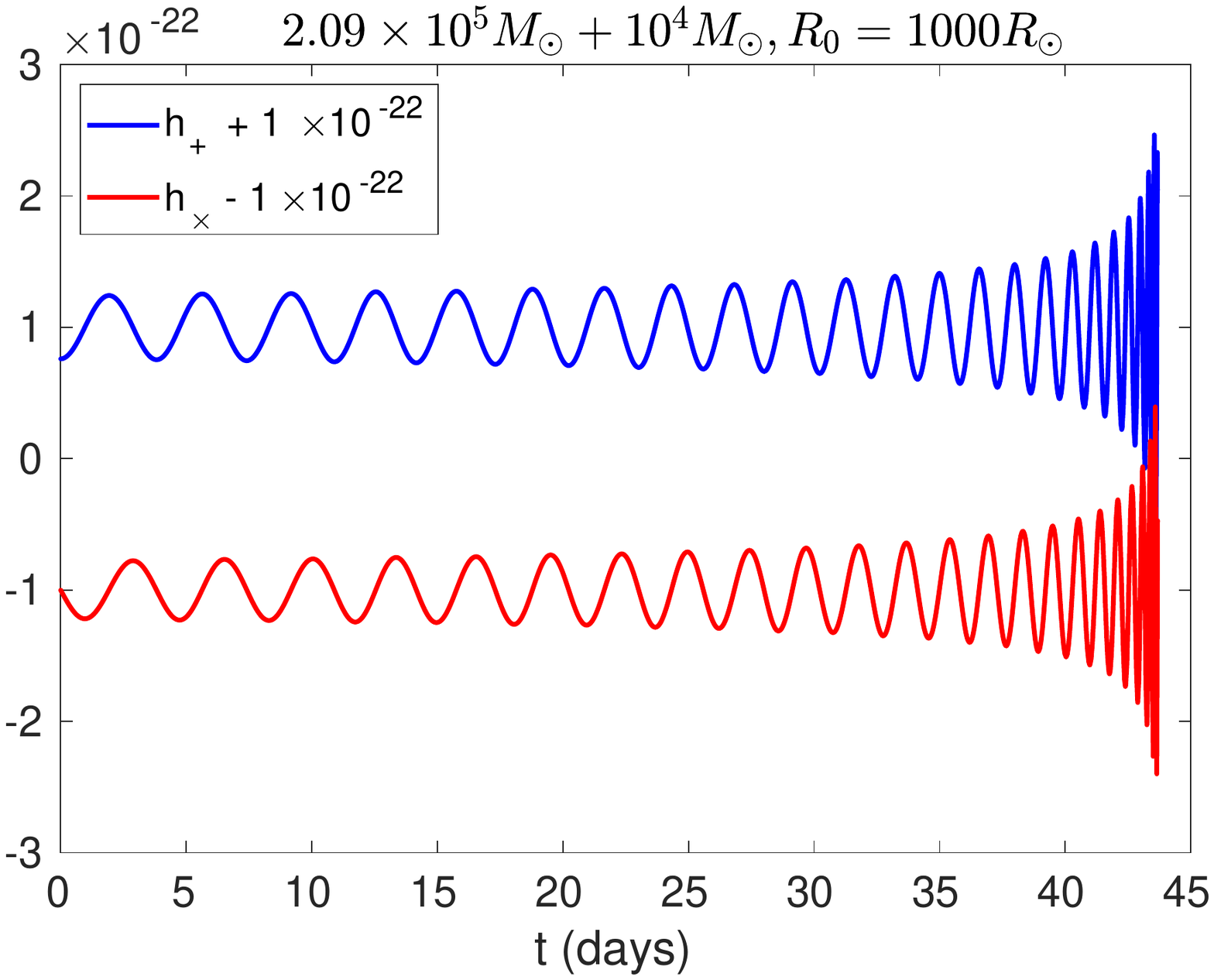} \includegraphics[scale=0.243,trim={2.25cm 6.65cm 1.45cm 5.9cm},clip]{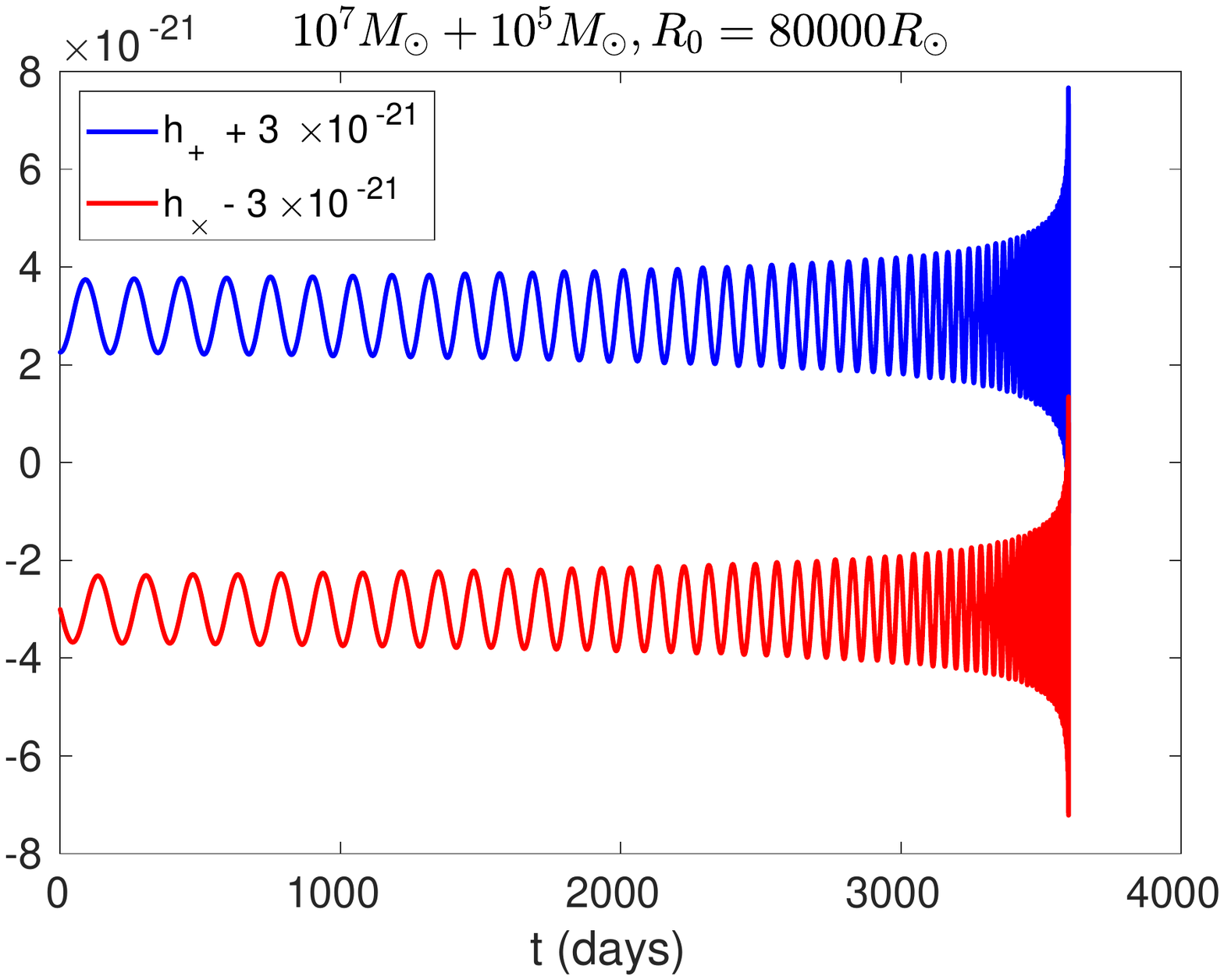}
\includegraphics[scale=0.232,trim={2.15cm 5.7cm 1.5cm 5.0cm},clip]{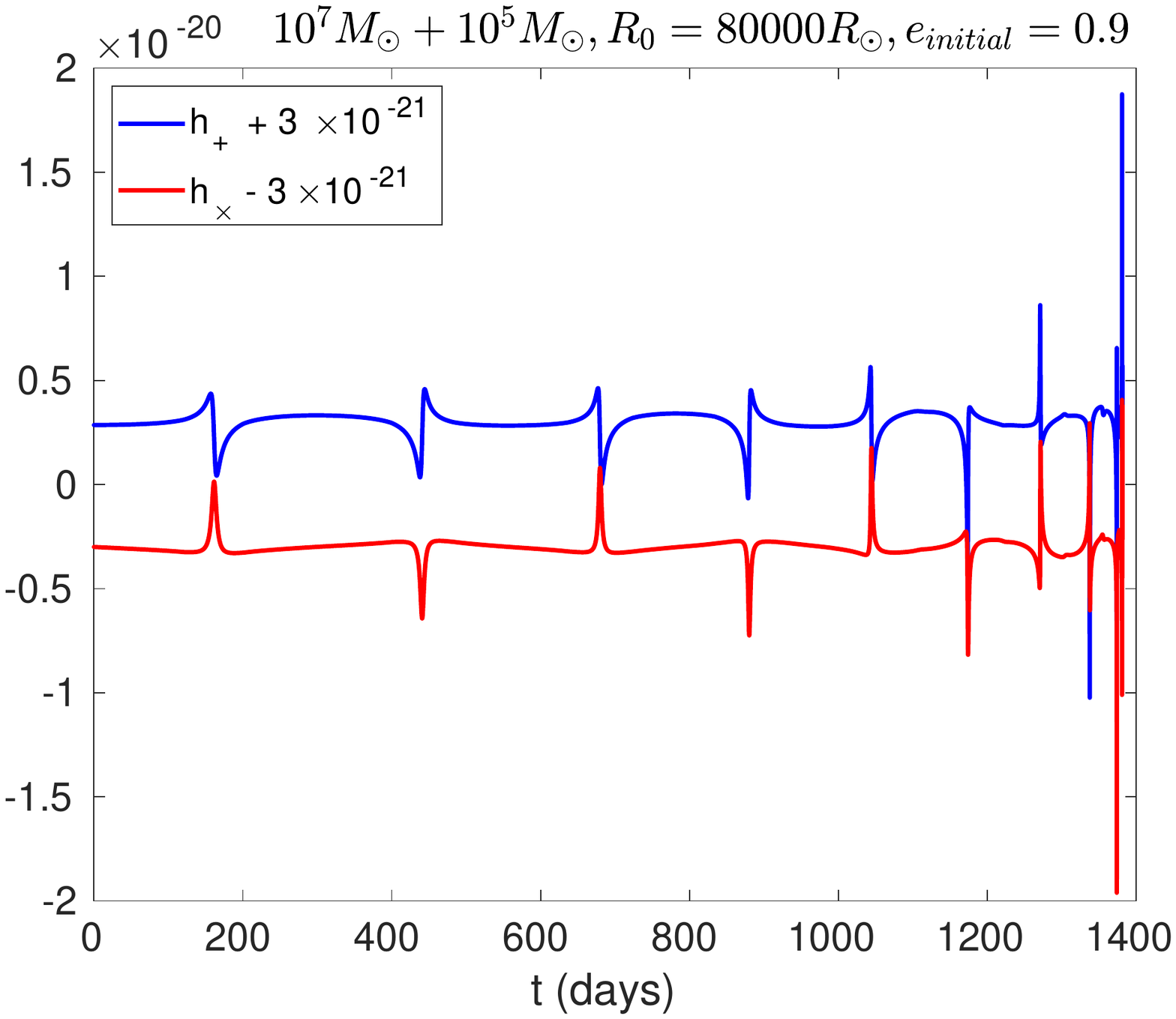}
\caption{Gravitational waveforms shown for binaries of companions of different masses and the SMSs' core, of approximately $6\times 10^4 M\ms$ for the $2.09\times 10^5 M\ms$ SMS and $10^4 M\ms$ for the $10^7 M\ms$ SMS, computed at redshift $z=6$. 
} \label{fig:WF}
\end{figure}
\paragraph*{}Recent studies suggest the formation of binary and even small multiples of SMSs in low-spin and high-spin halos, respectively \citep{2020ApJ...892L...4L,2020arXiv201211612P}. In \cite{2020ApJ...892L...4L} the authors present simulations of halos with accretion disks as massive as a few $10^5 M\ms$ with a few $10^3 - 10^4 M\ms$ fragments in the high-spin halos. With these studies in mind, and for each of the aforementioned SMS models we calculated the emitted waves, with $m=10-10^5 M\ms$ representing stellar to intermediate-mass BHs or other companions. The detector is assumed to be at a distance of $z = 6$\footnote{The system was taken to be at few $\textrm{Gpc}$ to $z = 6$, since supermassive primordial stars are suspected to be the progenitors of such quasars.}. The resulting wave-forms are shown in Fig.~\ref{fig:WF}. All in-spirals show a characteristic evolution beginning with regular low-amplitude oscillations at the early phases, which then gradually increase in frequency and amplitude down to the final plunge accompanied by a high amplitude burst.

\begin{figure}
\includegraphics[scale=0.455,trim={0.45cm 5.85cm 1.55cm 6.0cm},clip]{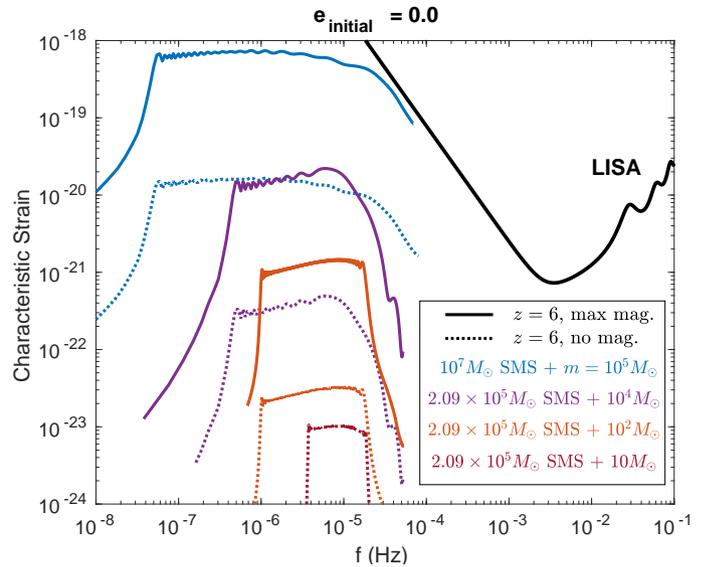}
\caption{Characteristic strains of the considered case studies, assuming $e_{initial}=0.0$, together with LISA sensitivity curve in black. The solid (dotted) lines correspond to $z=6$ with maximum magnification due to lensing (with no magnification due to lensing). 
} \label{fig:sms}
\end{figure}

\begin{figure}
\includegraphics[scale=0.455,trim={0.45cm 5.85cm 1.55cm 6.0cm},clip]{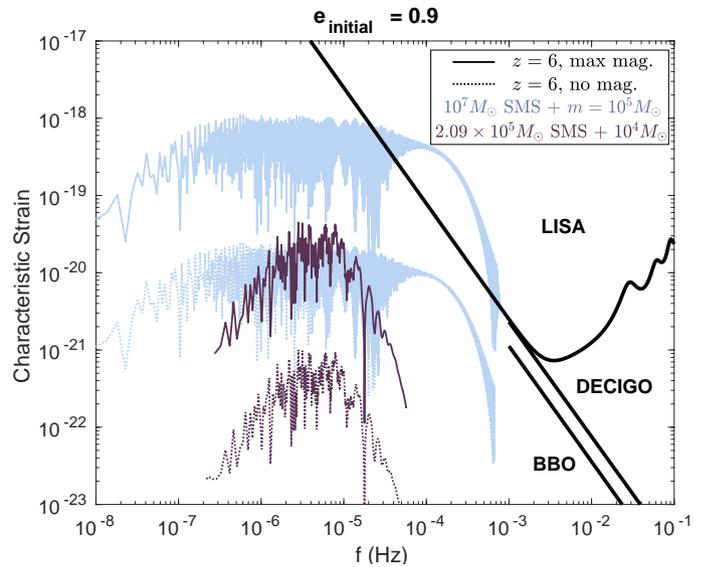}
\caption{Same as Fig.~\ref{fig:sms}, except $e_{initial}=0.9$.
} \label{fig:high_ecc}
\end{figure}

In Fig.~\ref{fig:sms} we show detection prospects of the GW signal from inspirals of companions of varied mass inside our SMS models. For simplicity we assumed a radiation-pressure-dominated gas with the adiabatic exponent\footnote{In a radiation-pressure-dominated gas
, gas pressure is only a small perturbation to the total pressure, and the total adiabatic exponent can be approximated as $\Gamma_1\approx 4/3+\beta/6$, where $\beta\approx 4 k_B/(\mu s)$, $\mu$ is the mean molecular weight, and $s/k_B$ is the entropy per baryon \citep{1939isss.book.....C}.} $\Gamma_1 = 4/3+0.009/6$ for the $2.09\times 10^5 M\ms$ SMS and $\Gamma_1 = 4/3$ for the $10^7 M\ms$ SMS. 
The calculated SNR for these case studies are low, SNR$=\mathcal{O}(0.01)-\mathcal{O}(10^{-4})$. In principle, such inspirals would have larger SNR and could be observable in the local universe, however, SMS were likely to form only at high redshifts, and therefore detections of inspirals in SMSs would require better sensitivity than expected in the upcoming GW-detector, or future GW detectors attuned to lower frequencies. 
Captured objects inside an SMS might initially follow eccentric orbits. The GW characteristic strain of the example eccentric orbit gets its peak at higher frequencies as seen in Fig.~\ref{fig:high_ecc}. That improves the SNR for these systems by a factor of $10-10^3$. 
Even such sources are not expected to be observed with currently planned missions. However, as we discuss below, high-redshift sources could be amplified through gravitational-lensing, and become observable, though only given favorable conditions,  

\section{Discussion} \label{sec:discussion}
The inspiral of massive stellar objects inside SMSs is driven by the interaction with the gaseous environment, and gives rise to unique GW-sources. However, SMSs are thought to have formed and evolved only at high redshifts \citep[z$>$6, see e.g.,][]{Habouzit2016}. The combined effect of red-shifting on the GW frequency and the decreased amplitude due to the large distances involved, preclude the potential detectability of such sources in currently planned GW-detectors. In the following we briefly discuss potential avenues that may allow the detection of GW sources from SMSs.

\subsection{Gravitational lensing}
Gravitational-lensing can amplify a faint GW signal. Although at low redshifts lensing would be rare, the probability for lensing is greatly increased at high redshifts \cite[e.g.][]{Pre+73}. The very high redshifts anticipated for the GW sources we discuss here raises the likelihood of gravitational-lensing as a significant factor in determining detection prospects. Although typical galaxy-cluster gravitational lenses can magnify background sources by a total magnification factor, $\lambda$, of up to $\lambda=50$, more extreme cases can give rise to magnifications of up to $\lambda=2000$  \citep{Kel+18}. High redshift GW sources as those discussed here could therefore potentially be significantly magnified through gravitational-lensing. The GW strain is proportional to $\sqrt{\lambda}$ and could therefore be increased by factors of up to $\sim45$ in the most extreme cases. As can be seen in Fig. \ref{fig:sms}, and more surely in Fig. \ref{fig:high_ecc}, such increase could potentially make the loudest sources we discuss be potentially observable, though possibly marginally, in LISA.
In particular, we calculate an SNR$=23$ for our most massive and initially eccentric inspiral. Such SNR makes it detectable with LISA even if the amplification is 4-5 lower (i.e. the lensing magnification is 16-25 times lower, at the level of $\sim 80-125$). Alternatively, at the highest magnification, even somewhat lower eccentricity inspirals could be detectable. 

\subsection{Supermassive quasi-stars}
Closely related to supermassive stars are the proposed ``quasi-stars'' \citep{Begelman08}, in which a rapidly-accreting BH is embedded within a massive, extended, radiation-dominated envelope. While the masses of any such objects would be comparable to SMSs \citep[and indeed, quasi-stars may arise from the collapse of supermassive stellar cores under some conditions, see][]{Begelman10}, the much more compact cores of quasi-stars would allow for GW radiation at much higher frequencies, likely allowing for their detection by the forthcoming LISA. Outside of the inner BH's Bondi radius, the structure of a quasi-star's envelope would be comparable to that of a SMS \citep{BT11}, and the motion of any merging stellar-mass fragment or compact object should proceed in a manner analogous to the cases described above. Within the Bondi radius, however, any such hydrostatic structure would be impossible. Given the uncertainties in modelling the density profile within this innermost region, in which the frequency of the GW radiation will peak, we reserve further study of mergers with quasi-stars for a subsequent work.

\subsection{Embedded stellar binaries}
Here we studied the inspirals of single objects on to the SMS core. However, compact binaries might also be captured/formed and become embedded inside SMSs. In such a case, the inner orbit of such binaries could also be affected by the interaction with the gas, which can drive them into merger, irrespective of the inspiral of the binary-system on the core of the SMS. Such a scenario could be somewhat similar to the scenario we suggested for black-hole binary mergers in AGN disks \citep{Mck+12} , driven by the interaction with the gas. These could therefore produce GW sources from mergers of binary SBHs/NSs, with effective frequencies in the LIGO band. However, these would be highly red-shifted given that SMSs arise only in the early Universe, potentially driving them outside the aLIGO band. Nevertheless, these would then be potentially observable by DECIGO.

\section{Summary} \label{sec:sum} 
In this study we explored the possibility of the GW-detection of inspirals of massive objects (stars or compact objects) inside the gaseous halos of supermassive stars. Such supermassive stars are thought to form in the early universe and later collapse to form intermediate mass black holes, which might be seeds that later grow and form supermassive black holes. The potential detection of such inspirals could provide a unique channel to study the structure of SMSs, and the origin of SMBH seeds. 

We follow the inspiral of stars inside the SMS due to the gravitational interaction of stars with the SMS large-scale atmosphere through gas-dynamical friction. The interaction decelerate the stars in the envelope and drive the emission of GW from the system. We characterize the GW signature from the inspiral and its detectability. We find that although the GW amplitude is sufficiently high as to be detected with upcoming space-based LISA GW detector in the local universe, the SNR from such sources is too low to be detected at high redshifts in which SMSs are thought to have formed. Consequently, future, higher sensitivity detectors will be required to identify such sources. Nevertheless, we also note that the understanding of SMS structure and evolution is still in its infancy, and improved models might give rise to different structures which would also affect the inspiral and GW characteristics. Moreover, we propose several channels producing potentially observable GW sources, if the high-redshift sources are magnified by gravitational-lensing. We also note, that mergers of stellar-binaries embedded in SMSs could also prove to be observable GW source, potentially observable in DECIGO, given their red-shifted frequencies.       


\section*{Acknowledgements}

T.E.W.\ acknowledges support from the National Research Council of Canada's Plaskett Fellowship. Y.B.G. acknowledges support from the Israeli Academy of Sciences' Adams fellowship.


\bibliographystyle{mnras}
\bibliography{seeds}

\begin{thebibliography}{}
\makeatletter
\relax
\def\mn@urlcharsother{\let\do\@makeother \do\$\do\&\do\#\do\^\do\_\do\%\do\~}
\def\mn@doi{\begingroup\mn@urlcharsother \@ifnextchar [ {\mn@doi@}
  {\mn@doi@[]}}
\def\mn@doi@[#1]#2{\def\@tempa{#1}\ifx\@tempa\@empty \href
  {http://dx.doi.org/#2} {doi:#2}\else \href {http://dx.doi.org/#2} {#1}\fi
  \endgroup}
\def\mn@eprint#1#2{\mn@eprint@#1:#2::\@nil}
\def\mn@eprint@arXiv#1{\href {http://arxiv.org/abs/#1} {{\tt arXiv:#1}}}
\def\mn@eprint@dblp#1{\href {http://dblp.uni-trier.de/rec/bibtex/#1.xml}
  {dblp:#1}}
\def\mn@eprint@#1:#2:#3:#4\@nil{\def\@tempa {#1}\def\@tempb {#2}\def\@tempc
  {#3}\ifx \@tempc \@empty \let \@tempc \@tempb \let \@tempb \@tempa \fi \ifx
  \@tempb \@empty \def\@tempb {arXiv}\fi \@ifundefined
  {mn@eprint@\@tempb}{\@tempb:\@tempc}{\expandafter \expandafter \csname
  mn@eprint@\@tempb\endcsname \expandafter{\@tempc}}}

\bibitem[\protect\citeauthoryear{{Abbott} et~al.,}{{Abbott}
  et~al.}{2016}]{2016PhRvD..93l2003A}
{Abbott} B.~P.,  et~al., 2016, \mn@doi [\prd] {10.1103/PhysRevD.93.122003},
  \href {https://ui.adsabs.harvard.edu/abs/2016PhRvD..93l2003A} {93, 122003}

\bibitem[\protect\citeauthoryear{{Aghanim} et~al.,}{{Aghanim}
  et~al.}{2018}]{Planck2018}
{Aghanim} N.,  et~al., 2018, arXiv e-prints, \href
  {https://ui.adsabs.harvard.edu/abs/2018arXiv180706209P} {p. arXiv:1807.06209}

\bibitem[\protect\citeauthoryear{{Babak} et~al.,}{{Babak}
  et~al.}{2017}]{2017PhRvD..95j3012B}
{Babak} S.,  et~al., 2017, \mn@doi [\prd] {10.1103/PhysRevD.95.103012}, \href
  {https://ui.adsabs.harvard.edu/abs/2017PhRvD..95j3012B} {95, 103012}

\bibitem[\protect\citeauthoryear{{Ball}, {Tout}, {{\.Z}ytkow}  \&
  {Eldridge}}{{Ball} et~al.}{2011}]{BT11}
{Ball} W.~H.,  {Tout} C.~A.,  {{\.Z}ytkow} A.~N.,   {Eldridge} J.~J.,  2011,
  \mn@doi [\mnras] {10.1111/j.1365-2966.2011.18591.x}, \href
  {https://ui.adsabs.harvard.edu/abs/2011MNRAS.414.2751B} {414, 2751}

\bibitem[\protect\citeauthoryear{{Begelman}}{{Begelman}}{2010}]{Begelman10}
{Begelman} M.~C.,  2010, \mn@doi [\mnras] {10.1111/j.1365-2966.2009.15916.x},
  \href {https://ui.adsabs.harvard.edu/abs/2010MNRAS.402..673B} {402, 673}

\bibitem[\protect\citeauthoryear{{Begelman}, {Rossi}  \& {Armitage}}{{Begelman}
  et~al.}{2008}]{Begelman08}
{Begelman} M.~C.,  {Rossi} E.~M.,   {Armitage} P.~J.,  2008, \mn@doi [\mnras]
  {10.1111/j.1365-2966.2008.13344.x}, \href
  {https://ui.adsabs.harvard.edu/abs/2008MNRAS.387.1649B} {387, 1649}

\bibitem[\protect\citeauthoryear{{Binney} \& {Tremaine}}{{Binney} \&
  {Tremaine}}{2008}]{binney2011galactic}
{Binney} J.,  {Tremaine} S.,  2008, {Galactic Dynamics: Second Edition}

\bibitem[\protect\citeauthoryear{{Chandrasekhar}}{{Chandrasekhar}}{1939}]{1939isss.book.....C}
{Chandrasekhar} S.,  1939, {An introduction to the study of stellar structure}

\bibitem[\protect\citeauthoryear{{Chon}, {Hosokawa}  \& {Yoshida}}{{Chon}
  et~al.}{2018}]{2018MNRAS.475.4104C}
{Chon} S.,  {Hosokawa} T.,   {Yoshida} N.,  2018, \mn@doi [\mnras]
  {10.1093/mnras/sty086}, \href
  {https://ui.adsabs.harvard.edu/abs/2018MNRAS.475.4104C} {475, 4104}

\bibitem[\protect\citeauthoryear{{Dijkstra} et~al.}{{Dijkstra}
  et~al.}{2008}]{2008MNRAS.391.1961D}
{Dijkstra} M.,  et~al., 2008, \mn@doi [\mnras]
  {10.1111/j.1365-2966.2008.14031.x}, \href
  {https://ui.adsabs.harvard.edu/abs/2008MNRAS.391.1961D} {391, 1961}

\bibitem[\protect\citeauthoryear{{Ferrarese} \& {Ford}}{{Ferrarese} \&
  {Ford}}{2005}]{2005SSRv..116..523F}
{Ferrarese} L.,  {Ford} H.,  2005, \mn@doi [\ssr] {10.1007/s11214-005-3947-6},
  \href {https://ui.adsabs.harvard.edu/abs/2005SSRv..116..523F} {116, 523}

\bibitem[\protect\citeauthoryear{{Ginat}, {Glanz}, {Perets}, {Grishin}  \&
  {Desjacques}}{{Ginat} et~al.}{2020}]{2019arXiv190311072G}
{Ginat} Y.~B.,  {Glanz} H.,  {Perets} H.~B.,  {Grishin} E.,   {Desjacques} V.,
  2020, \mn@doi [\mnras] {10.1093/mnras/staa465}, \href
  {https://ui.adsabs.harvard.edu/abs/2020MNRAS.493.4861G} {493, 4861}

\bibitem[\protect\citeauthoryear{{Habouzit}, {Volonteri}, {Latif}, {Dubois}  \&
  {Peirani}}{{Habouzit} et~al.}{2016}]{Habouzit2016}
{Habouzit} M.,  {Volonteri} M.,  {Latif} M.,  {Dubois} Y.,   {Peirani} S.,
  2016, \mn@doi [\mnras] {10.1093/mnras/stw1924}, \href
  {https://ui.adsabs.harvard.edu/abs/2016MNRAS.463..529H} {463, 529}

\bibitem[\protect\citeauthoryear{{Haemmerl{\'e}}}{{Haemmerl{\'e}}}{2020a}]{2020arXiv201008229H}
{Haemmerl{\'e}} L.,  2020a, arXiv e-prints, \href
  {https://ui.adsabs.harvard.edu/abs/2020arXiv201008229H} {p. arXiv:2010.08229}

\bibitem[\protect\citeauthoryear{{Haemmerl{\'e}}}{{Haemmerl{\'e}}}{2020b}]{2020A&A...644A.154H}
{Haemmerl{\'e}} L.,  2020b, \mn@doi [\aap] {10.1051/0004-6361/202039828}, \href
  {https://ui.adsabs.harvard.edu/abs/2020A&A...644A.154H} {644, A154}

\bibitem[\protect\citeauthoryear{{Haemmerl{\'e}}, {Woods}, {Klessen}, {Heger}
  \& {Whalen}}{{Haemmerl{\'e}} et~al.}{2018}]{2018MNRAS.474.2757H}
{Haemmerl{\'e}} L.,  {Woods} T.~E.,  {Klessen} R.~S.,  {Heger} A.,   {Whalen}
  D.~J.,  2018, \mn@doi [\mnras] {10.1093/mnras/stx2919}, \href
  {https://ui.adsabs.harvard.edu/abs/2018MNRAS.474.2757H} {474, 2757}

\bibitem[\protect\citeauthoryear{{Hartwig}, {Agarwal}  \& {Regan}}{{Hartwig}
  et~al.}{2018}]{2018MNRAS.479L..23H}
{Hartwig} T.,  {Agarwal} B.,   {Regan} J.~A.,  2018, \mn@doi [\mnras]
  {10.1093/mnrasl/sly091}, \href
  {https://ui.adsabs.harvard.edu/abs/2018MNRAS.479L..23H} {479, L23}

\bibitem[\protect\citeauthoryear{{Hosokawa}, {Yorke}, {Inayoshi}, {Omukai}  \&
  {Yoshida}}{{Hosokawa} et~al.}{2013}]{2013ApJ...778..178H}
{Hosokawa} T.,  {Yorke} H.~W.,  {Inayoshi} K.,  {Omukai} K.,   {Yoshida} N.,
  2013, \mn@doi [\apj] {10.1088/0004-637X/778/2/178}, \href
  {https://ui.adsabs.harvard.edu/abs/2013ApJ...778..178H} {778, 178}

\bibitem[\protect\citeauthoryear{{Inayoshi}, {Visbal}  \& {Haiman}}{{Inayoshi}
  et~al.}{2020}]{2020ARA&A..58...27I}
{Inayoshi} K.,  {Visbal} E.,   {Haiman} Z.,  2020, \mn@doi [\araa]
  {10.1146/annurev-astro-120419-014455}, \href
  {https://ui.adsabs.harvard.edu/abs/2020ARA&A..58...27I} {58, 27}

\bibitem[\protect\citeauthoryear{{Kelly} et~al.,}{{Kelly}
  et~al.}{2018}]{Kel+18}
{Kelly} P.~L.,  et~al., 2018, \mn@doi [Nature Astronomy]
  {10.1038/s41550-018-0430-3}, \href
  {https://ui.adsabs.harvard.edu/abs/2018NatAs...2..334K} {2, 334}

\bibitem[\protect\citeauthoryear{{Kim} \& {Kim}}{{Kim} \&
  {Kim}}{2007}]{2007ApJ...665..432K}
{Kim} H.,  {Kim} W.-T.,  2007, \mn@doi [\apj] {10.1086/519302}, \href
  {https://ui.adsabs.harvard.edu/abs/2007ApJ...665..432K} {665, 432}

\bibitem[\protect\citeauthoryear{{Kormendy} \& {Ho}}{{Kormendy} \&
  {Ho}}{2013}]{2013ARA&A..51..511K}
{Kormendy} J.,  {Ho} L.~C.,  2013, \mn@doi [\araa]
  {10.1146/annurev-astro-082708-101811}, \href
  {https://ui.adsabs.harvard.edu/abs/2013ARA&A..51..511K} {51, 511}

\bibitem[\protect\citeauthoryear{{Latif}, {Khochfar}  \& {Whalen}}{{Latif}
  et~al.}{2020}]{2020ApJ...892L...4L}
{Latif} M.~A.,  {Khochfar} S.,   {Whalen} D.,  2020, \mn@doi [\apjl]
  {10.3847/2041-8213/ab7c61}, \href
  {https://ui.adsabs.harvard.edu/abs/2020ApJ...892L...4L} {892, L4}

\bibitem[\protect\citeauthoryear{{Lincoln} \& {Will}}{{Lincoln} \&
  {Will}}{1990}]{1990PhRvD..42.1123L}
{Lincoln} C.~W.,  {Will} C.~M.,  1990, \mn@doi [\prd]
  {10.1103/PhysRevD.42.1123}, \href
  {https://ui.adsabs.harvard.edu/abs/1990PhRvD..42.1123L} {42, 1123}

\bibitem[\protect\citeauthoryear{Maggiore}{Maggiore}{2008}]{maggiore2008gravitational}
Maggiore M.,  2008, Gravitational Waves: Volume 1: Theory and Experiments.
Gravitational Waves, OUP Oxford

\bibitem[\protect\citeauthoryear{{Mayer}, {Kazantzidis}, {Escala}  \&
  {Callegari}}{{Mayer} et~al.}{2010}]{2010Natur.466.1082M}
{Mayer} L.,  {Kazantzidis} S.,  {Escala} A.,   {Callegari} S.,  2010, \mn@doi
  [\nat] {10.1038/nature09294}, \href
  {https://ui.adsabs.harvard.edu/abs/2010Natur.466.1082M} {466, 1082}

\bibitem[\protect\citeauthoryear{{McKernan}, {Ford}, {Lyra}  \&
  {Perets}}{{McKernan} et~al.}{2012}]{Mck+12}
{McKernan} B.,  {Ford} K.~E.~S.,  {Lyra} W.,   {Perets} H.~B.,  2012, \mn@doi
  [\mnras] {10.1111/j.1365-2966.2012.21486.x}, \href
  {https://ui.adsabs.harvard.edu/abs/2012MNRAS.425..460M} {425, 460}

\bibitem[\protect\citeauthoryear{{Moore}, {Cole}  \& {Berry}}{{Moore}
  et~al.}{2015}]{2015CQGra..32a5014M}
{Moore} C.~J.,  {Cole} R.~H.,   {Berry} C.~P.~L.,  2015, CQGra, 32, 015014

\bibitem[\protect\citeauthoryear{{Ostriker}}{{Ostriker}}{1999}]{1999ApJ...513..252O}
{Ostriker} E.~C.,  1999, \mn@doi [\apj] {10.1086/306858}, \href
  {https://ui.adsabs.harvard.edu/abs/1999ApJ...513..252O} {513, 252}

\bibitem[\protect\citeauthoryear{{Patrick}, {Whalen}, {Elford}  \&
  {Latif}}{{Patrick} et~al.}{2020}]{2020arXiv201211612P}
{Patrick} S.~J.,  {Whalen} D.~J.,  {Elford} J.~S.,   {Latif} M.~A.,  2020,
  arXiv e-prints, \href {https://ui.adsabs.harvard.edu/abs/2020arXiv201211612P}
  {p. arXiv:2012.11612}

\bibitem[\protect\citeauthoryear{{Press} \& {Gunn}}{{Press} \&
  {Gunn}}{1973}]{Pre+73}
{Press} W.~H.,  {Gunn} J.~E.,  1973, \mn@doi [\apj] {10.1086/152430}, \href
  {https://ui.adsabs.harvard.edu/abs/1973ApJ...185..397P} {185, 397}

\bibitem[\protect\citeauthoryear{{Sakurai}, {Hosokawa}, {Yoshida}  \&
  {Yorke}}{{Sakurai} et~al.}{2015}]{2015MNRAS.452..755S}
{Sakurai} Y.,  {Hosokawa} T.,  {Yoshida} N.,   {Yorke} H.~W.,  2015, \mn@doi
  [\mnras] {10.1093/mnras/stv1346}, \href
  {https://ui.adsabs.harvard.edu/abs/2015MNRAS.452..755S} {452, 755}

\bibitem[\protect\citeauthoryear{{Sesana}, {Volonteri}  \& {Haardt}}{{Sesana}
  et~al.}{2007}]{2007MNRAS.377.1711S}
{Sesana} A.,  {Volonteri} M.,   {Haardt} F.,  2007, \mn@doi [\mnras]
  {10.1111/j.1365-2966.2007.11734.x}, \href
  {https://ui.adsabs.harvard.edu/abs/2007MNRAS.377.1711S} {377, 1711}

\bibitem[\protect\citeauthoryear{{Tagawa}, {Haiman}  \& {Kocsis}}{{Tagawa}
  et~al.}{2019}]{2019arXiv190910517T}
{Tagawa} H.,  {Haiman} Z.,   {Kocsis} B.,  2019, arXiv e-prints, \href
  {https://ui.adsabs.harvard.edu/abs/2019arXiv190910517T} {p. arXiv:1909.10517}

\bibitem[\protect\citeauthoryear{{Volonteri}}{{Volonteri}}{2010}]{2010A&ARv..18..279V}
{Volonteri} M.,  2010, \mn@doi [\aapr] {10.1007/s00159-010-0029-x}, \href
  {https://ui.adsabs.harvard.edu/abs/2010A&ARv..18..279V} {18, 279}

\bibitem[\protect\citeauthoryear{{Wise}, {Regan}, {O'Shea}, {Norman}, {Downes}
  \& {Xu}}{{Wise} et~al.}{2019}]{2019Natur.566...85W}
{Wise} J.~H.,  {Regan} J.~A.,  {O'Shea} B.~W.,  {Norman} M.~L.,  {Downes}
  T.~P.,   {Xu} H.,  2019, \mn@doi [\nat] {10.1038/s41586-019-0873-4}, \href
  {https://ui.adsabs.harvard.edu/abs/2019Natur.566...85W} {566, 85}

\bibitem[\protect\citeauthoryear{{Woods}, {Heger}, {Whalen}, {Haemmerl{\'e}}
  \& {Klessen}}{{Woods} et~al.}{2017}]{2017ApJ...842L...6W}
{Woods} T.~E.,  {Heger} A.,  {Whalen} D.~J.,  {Haemmerl{\'e}} L.,   {Klessen}
  R.~S.,  2017, \mn@doi [\apjl] {10.3847/2041-8213/aa7412}, \href
  {https://ui.adsabs.harvard.edu/abs/2017ApJ...842L...6W} {842, L6}

\bibitem[\protect\citeauthoryear{{Woods} et~al.,}{{Woods}
  et~al.}{2019}]{2019PASA...36...27W}
{Woods} T.~E.,  et~al., 2019, \mn@doi [\pasa] {10.1017/pasa.2019.14}, \href
  {https://ui.adsabs.harvard.edu/abs/2019PASA...36...27W} {36, e027}

\bibitem[\protect\citeauthoryear{{Woods}, {Patrick}, {Elford}, {Whalen}  \&
  {Heger}}{{Woods} et~al.}{2021}]{2021arXiv210208963W}
{Woods} T.~E.,  {Patrick} S.,  {Elford} J.~S.,  {Whalen} D.~J.,   {Heger} A.,
  2021, arXiv e-prints, \href
  {https://ui.adsabs.harvard.edu/abs/2021arXiv210208963W} {p. arXiv:2102.08963}

\bibitem[\protect\citeauthoryear{{Woosley}, {Heger}  \& {Weaver}}{{Woosley}
  et~al.}{2002}]{2002RvMP...74.1015W}
{Woosley} S.~E.,  {Heger} A.,   {Weaver} T.~A.,  2002, \mn@doi [Reviews of
  Modern Physics] {10.1103/RevModPhys.74.1015}, \href
  {https://ui.adsabs.harvard.edu/abs/2002RvMP...74.1015W} {74, 1015}

\makeatother
\end{thebibliography}





\bsp
\label{lastpage}
\end{document}